\documentclass[twocolumn,showpacs,preprintnumbers,amsmath,amssymb]{revtex4}
\usepackage{graphicx}% Include figure files
\usepackage{dcolumn}% Align table columns on decimal point
\usepackage{bm}% bold math

\begin{document}

%\preprint{DV/PDAC}

\title{How accurate is Poisson-Boltzmann theory for monovalent ions near highly charged interfaces?}

\author{Wei Bu, David Vaknin, and Alex Travesset}
\affiliation{Ames Laboratory and Department of Physics and Astronomy Iowa State
University, Ames, Iowa 50011, USA}
\date{\today --- Submitted to \emph{Langmuir} 10-Dec-2005}

\begin{abstract}
Surface sensitive synchrotron x-ray scattering studies were performed to obtain
the distribution of monovalent ions next to a highly charged interface. A lipid
phosphate (dihexadecyl hydrogen-phosphate) was spread as a monolayer at the
air-water interface to control surface charge density.  Using anomalous
reflectivity off and at the $L_3 $  Cs$^{+}$ resonance, we provide spatial
counterion (Cs$^{+}$) distributions next to the negatively charged interfaces.
Five decades in bulk concentrations are investigated, demonstrating that the
interfacial distribution is strongly dependent on bulk concentration.  We show
that this is due to the strong binding constant of hydronium H$_3$O$^{+}$ to
the phosphate group, leading to proton transfer back to the phosphate group and
to a reduced surface charge. The increase of Cs$^+$ concentration modifies the
contact value potential, thereby causing proton release.  This process
effectively modifies surface charge density and enables exploration of ion
distributions as a function of effective surface charge-density.  The
experimentally obtained ion distributions are compared to distributions
calculated by Poisson-Boltzmann theory accounting for the variation of surface
charge density due to proton release and binding.  We also discuss the accuracy
of our experimental results in discriminating among possible deviations from
Poisson-Boltzmann theory.
%\verb+\pacs{#1}+ command.
\end{abstract}

%\pacs{82.70.Uv,82.35.Rs,81.07.Nb}

%Use showkeys class option if keyword
\maketitle

\section{Introduction}\label{Introduction}

The theoretical determination of ion distributions in aqueous solutions was
initiated almost a century ago by Guoy \cite{Guoy1910} and Chapman
\cite{Chapman1913}, who applied the Poisson-Boltzmann (PB) theory to calculate
the spatial distribution of monovalent ions near a uniformly charged interface.
Ever since their seminal work, the topic remains central in statistical
mechanics, physical chemistry, and biophysics\cite{Israelachvili2000}.  The
original PB theory is a mean field theory with some simplified assumptions such
as, point like particles and uniform surface charge density.   To account for
the finite ionic radius, Stern introduced phenomenologically a layer of ions at
the charged interface with a different dielectric constant, the Stern layer
\cite{Stern1924}.  The effect of excess salt concentration and the resulting
screening was extended by Debye-H{\"uckel} \cite{Debye1923}. Grahame
generalized the Gouy-Chapman theory to multivalent ions \cite{Grahame1947}.
Subsequently, more refined theories and numerical simulations were developed to
incorporate short-range interactions, image charges, finite size ionic radius,
and ion-ion correlations
\cite{Outhwaitie1978,Outhwaitie1980,Torrie1982,Kjellander1984}. More recently,
modifications of PB theory have been developed to incorporate hydration forces
\cite{Leikin1993,Manciu2004,Faraudo2004}. Some first-principles calculations of
surface-tension for amphiphilic monolayers  assume a PB theory with one or more
layers of varying dielectric constant \cite{Proesser2001}.

Experimental support for the validity of PB theory was provided by
electrokinetic, visco-electric effects, and other techniques (see Ref.
\cite{Hunter1981} for a review).  McLaughlin and collaborators
\cite{McLaughlin1989} have shown good agreement between $\zeta$-potentials
computed from PB theories and electrophoretic measurements in lipid vesicles.
Other techniques, such as radiotracer experiments
\cite{Tajima1970,Tajima1971,Kobayashi1986},  x-ray reflectivity
\cite{Kjaer1989,Bloch1990,Bedzyk1990}, or infrared spectroscopy
\cite{LeCalvez2001} allow the determination of the total amount of ions in the
immediate vicinity of a charged interface.  It is noteworthy that all the
experimental data for monovalent ions (at moderate salt concentrations
$\lesssim$ 0.1 M) outlined above are adequately described by the Guoy-Chapman
theory (with the generalization of excess salt) with no need for further
corrections \cite{Bloch1990,Travesset2006}.  A close inspection, however, shows
the agreement between theory and experiment is either based on fitting
variables such as surface charge or interfacial dielectric values, that are not
known in advance or/and based on integrated quantities.   As an example, it has
been shown recently that the degree of proton dissociation of arachidic acid
spread on a sodium salt solution is adequately described by PB theory, but this
agreement only involves the integral (over the entire space) of the sodium
distribution \cite{LeCalvez2001,Travesset2006}.  Thus, local deviations that
preserve the integral of the distribution (i.e., total number of ions) are not
discriminated by these experiments.  On the other hand, force measurements
between two charged membranes separated by salt solutions, although well
described by PB theory at large distances, show strong deviations at short
distances (1-2 nm)\cite{Leikin1993,Israelachvili2000}.  The origin of these
hydration forces is still controversial.  In some cases, it has been suggested
to extend PB theory to incorporate the restructuring of water, resulting in ion
distributions that deviate from PB at short distances from the interface
\cite{Manciu2004,Gur1977,Paunov1996}.  It is therefore imperative to determine
the ion distribution itself to establish the degree of accuracy of PB theory.

In this manuscript, we experimentally determine the distribution of Cs$^+$ ions
next to charged interfaces by anomalous x-ray reflectivity
\cite{Vaknin2003,Bu2005} for several decades in bulk cesium concentrations. We
compare our results with the predictions of PB theory and present a discussion
on the sensitivity of the experimental results to quantify the magnitude of
possible deviations from the PB distribution.  This kind of investigation has
become feasible only with the advent of the second generation x-ray synchrotron
sources with novel insertion devices (i.e. undulator) and improved optics,
which readily produce variable-energy x-ray beams with brilliances capable of
detecting a single atomic-layer even if not closely packed.

The manuscript is organized as follows: In Section~\ref{Theory}, we review
several theoretical results relevant to this study. In
Section~\ref{Experimntal}, the experimental details are provided.  In
Section~\ref{Results}, the x-ray reflectivity (XR) and grazing incidence x-ray
diffraction (GIXD) measurements are presented with detailed structural
analysis. The experimental results are compared with the theoretical PB theory
in Section~\ref{Analysis}.  The implications of the present study are discussed
in Section~\ref{Conclusions}.

\section{PB theory including proton transfer and release}\label{Theory}

\begin{figure}[!]
\includegraphics[width=3.2 in]{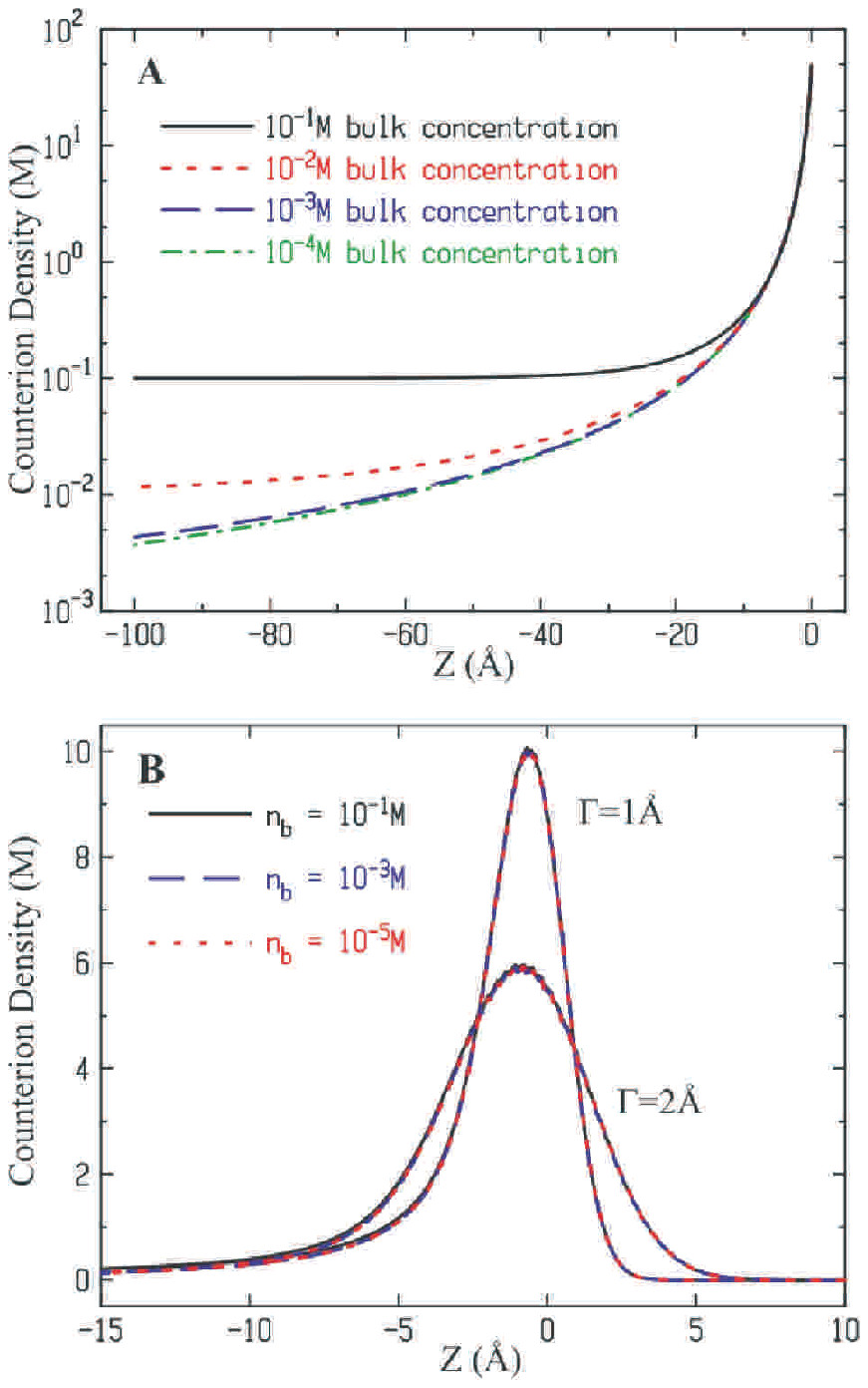}
\caption{\label{pb} (A) Calculated monovalent ion distributions $n^{+}(z)$ near
a negatively charged surface as obtained from PB theory, Eq.\ (\ref{n_z}) for
different bulk concentration.  Surface charge density $\sigma_0$ is one
electron charge per 40 {\AA}$^{2}$, $\epsilon=80$, and $T = 292$ K. Note that
for this surface charge density the value of the distribution at the $z=0$ is
practically a constant, $n^{+}(0) \approx \frac{2\pi\sigma_0^2}{k_BT\epsilon}$
(B) The convolutions of the distributions $n^{+}(z)$ using Eq.\
(\ref{Conv_Gauss}) assuming two $\Gamma$ values as indicated. For this strongly
charged surface the convoluted distributions for different bulk ionic
concentrations are practically indistinguishable.}
\end{figure}

We briefly review the PB theory of an aqueous solution containing monovalent
ions (1:1 positive and negative electrolytes) of average bulk concentration
$n_b$ in the presence of an ideally flat charged interface of known surface
density $\sigma_0 = -en_0$, at $z = 0$.  As the surface charge is uniformly
distributed, the electric potential $\psi$ depends only on the distance from
the interface $z$, and the Boltzmann distribution is
\begin{equation}
\rho(z)=en_{b}[\mbox{e}^{-e\psi(z)/k_BT}-\mbox{e}^{e\psi(z)/k_BT}]=-2en_{b}\sinh{\phi(z)},
\end{equation}
where $k_B$ is the Boltzmann constant and $\phi(z)= e\psi(z)/k_BT$. The ion
distribution is calculated self-consistently from the Poisson equation
\begin{equation}
d^{2}\phi/dz^{2}=\sinh\phi/\lambda^{2}_{D} \label{PB1}
\end{equation}
where $\lambda_{D}=(\epsilon k_{B}T/8\pi e^{2}n_{b})^{1/2}$ is the Debye
screening length.  Equation\ (\ref{PB1}) can be solved analytically, yielding
\begin{equation}
\psi(z)=-\frac{2k_{B}T}{e}\ln\left[\frac{1+\gamma e^{-z/\lambda_{D}}}{1-\gamma
e^{-z/\lambda_{D}}}\right], \label{psi_z}
\end{equation}
\begin{equation}
n^{+}(z)=n_{b}\left(\frac{1+\gamma e^{-z/\lambda_{D}}}{1-\gamma
e^{-z/\lambda_{D}}}\right)^{2}, \label{n_z}
\end{equation}
where
$\gamma=\tanh[e\psi(0)/4k_{B}T]$=$-\lambda_{GC}/\lambda_{D}+((\lambda_{GC}/\lambda_{D})^{2}+1)^{1/2}$,
and $\lambda_{GC}=k_{B}T\epsilon/2\pi\sigma_{0}e$ is the Gouy-Chapman length.
Figure\ \ref{pb}(A) shows calculated counter-ion distributions using Eq.\
(\ref{n_z}) for a fixed surface charge density, typical of fully deprotonated
closed packed phospholipids $\sigma_0 = -e/40$ {\AA}$^2$ for several salt
concentrations.  It is worth noting that for this surface density
($\lambda_{GC} \approx 0.9$ {\AA}) and bulk salt concentrations ($\lambda_D =
963.5$ {\AA} and 9.6 {\AA} at 10$^{-5}$ and 0.1 M, respectively)
$\frac{\lambda_{GC}}{\lambda_D}<< 1$, and $\gamma \approx
1-\frac{\lambda_{GC}}{\lambda_D}$, the concentration of ions next to the
interface is independent of bulk concentration
\begin{equation}\label{GC_dist}
    n(z)\approx \frac{\epsilon k_BT}{2\pi e^2  (z+\lambda_{GC})^2};\ \textrm{ for} \frac{z}{\lambda_{D}}\ll 1.
\end{equation}

We have recently argued \cite{Bu2005} that theoretical PB distributions should
be convoluted with the experimental resolution function $R(z)$ for comparison
with experiments, as follows,
\begin{equation} n_r^+(z) =
\int{n^+(z^{\prime})R(z^{\prime}-z)\text{dz}^{\prime}}. \label{Conv_Gen}
\end{equation}
In the particular case of a gaussian resolution function the convolution is
\begin{equation} n_r^+(z) =
\frac{1}{\Gamma\sqrt{2\pi}}\int{n^+(z^{\prime})e^{-\frac{(z-z^{\prime})^2}{2\Gamma^2}}\text{dz}^{\prime}}.
\label{Conv_Gauss}
\end{equation}
 The width in the Gaussian function is given as a sum, $\Gamma=\sqrt{\sum\Gamma_i^2}$, where each
$\Gamma_i$'s account for surface-roughness due to capillary waves and
imperfections in the monolayer and  other contributions. Since
surface-roughness is the dominant contribution we use $\Gamma\approx\xi_{Ref}$,
where $\xi_{Ref}$ is the average roughness determined from the reflectivities.
From the statistical mechanical perspective, the above convolution maybe
justified if the fluctuations of the interface are gaussian and the wavelength
of the capillary waves $\Lambda$ is large ($\Lambda \gg \delta$, where $\delta$
is typical ion size) so that the ions ``see'' an effective flat charged
interface. Alternative methods for computing distributions next to flexible
interfaces have recently been proposed \cite{Fleck2005}.  Figure\ \ref{pb}(B)
shows convolution of the distributions shown in Fig.\ \ref{pb}(A) for two
different values of $\Gamma$ as indicated. It is interesting to note that the
convoluted calculations are practically indistinguishable as a function of bulk
salt concentration. In Section\ \ref{Analysis} we discuss quantitatively to
what extent the experimental resolution limits our ability to compare with
theoretical distributions.

The system we describe in this paper differs from simple PB in that one of the
ions species, the hydronium H$_3$O$^{+}$, can bind to the interface, and as a
result, out of the $N_P$ phosphate groups forming the interface, only $\alpha
N_P$ are electrically charged.  The density of charge in the system (including
the interfacial charge $z=0$) is therefore
\begin{equation}\label{Two_State}
\rho(z)=e \sum_{a} q_a n_a(z)-e \alpha n_0 \delta(z).
\end{equation}
The index $a$ runs over the different species of ions present in
solution and $q_a=\pm 1$ is their valence (i.e. the solution
contains monovalent ions only). In the free energy, the ions are
treated as an ideal gas within the solvent, and subjected to
electrostatic contributions (the standard PB) \cite{Safran1994}.
We also incorporate two additional terms that represent the
favorable binding of hydroniums to the interface and the mixing
entropy of the charged and neutral phosphate groups at the
interface,
\begin{eqnarray}\label{Free_Energy}
  F &=& \frac{1}{2} \int d^3 {\bf r} d^3 {\bf r^{\prime}} \rho({\bf r})
\frac{1}{|{\bf r}-{\bf r^{\prime}}|} \rho({\bf r}^{\prime})
\\\nonumber
   &+&k_BT \sum_{a} \int d^3{\bf r} n_a({\bf r})(\log(v n_a({\bf
   r}))-1)\\\nonumber
   &-&E_H(1-\alpha)N_P-k_BT \log\left(\frac{N_P!}{(\alpha N_P
   )!((1-\alpha)N_P)!}\right) \ ,
\end{eqnarray}
where $E_H$ is the gain in free energy for a hydronium binding to
the phosphate group. The quantity $v$ has dimensions of volume and
defines the standard state.

The condition that the free energy is a minimum for variations of the different
number densities $n_a({\bf r})$ under the constraint of a fixed number of
particles (canonical ensemble) leads to the Boltzmann distribution
$n_a(z)=v^{-1} e^{\frac{\mu_a}{k_B T}} e^{-q_a \frac{e \psi(z)}{k_B T}}$ with
chemical potential $\mu_a=k_B T \log( n_a^b v)$, and also to the Poisson
equation Eq.\ (\ref{PB1}). Further minimization of the free energy with respect
to the parameter $\alpha$ yields the additional equation
\begin{equation}\label{Energy}
    E_H/k_BT-\phi(0)+\log(\alpha N_P)-\log((1-\alpha)N_P)=-\mu_H/k_BT
\end{equation}
where $\psi(0)$ is the value of the electric potential at the
interface (the contact value). This equation may be rewritten as
\begin{equation}\label{Chemical_Binding}
    n_H^{b} v e^{-\phi(0)}\frac{\alpha}{1-\alpha}= e^{-\frac{E_H}{k_B T}} \ ,
\end{equation}
and expresses the equilibrium of the process H$_3$O$^{+}+$PO$_4^{-}$
 $\longleftrightarrow$ PO$_4$H$^{0}+$H$_2$O \cite{Healy1978}. If the standard
state is defined as $v= 1$ molar, the binding free energy $E_H$ is related to
the $pK_a$ of the molecules forming the interface by $E_H=k_BT\log K_a$ and the
hydronium concentration is related to the $pH$ of the solution by
n$_H=10^{-pH}$. Substitution of these values into Eq.\ (\ref{Chemical_Binding})
yields
\begin{equation}\label{pH_eq}
    10^{-pH}  e^{-\phi(0)}\frac{\alpha}{1-\alpha}= 10^{-pK_a} \ .
\end{equation}
The fraction of charged phosphate groups is then
\begin{equation}\label{alpha_G}
\alpha=\frac{1}{1+10^{-(pH-pK_a)}e^{-\phi(0)}}.
\end{equation}
As a result, the PB equation Eq.\ (\ref{PB1}) needs to be solved with the
self-consistent boundary condition Eq.\ (\ref{alpha_G})
\begin{equation}\label{Self_Consistent}
    \sinh\left(\frac{\phi(0)}{2}\right)=-\left(\frac{\lambda_D}{\lambda_{GC}}\right)
    \frac{1}{1+10^{-(pH-pK_a)}e^{-\phi(0)}}.
\end{equation}
 The ion distribution is given by Eq.\ (\ref{n_z}) with a renormalized Guoy-Chapman length
$\lambda_{GC}^{\prime}=\lambda_{GC}/\alpha$ (RPB). At high surface charge, the
contact point potential is significantly larger than $k_BT$, and the
consequences of Eq.\ (\ref{Self_Consistent}) are dramatic. The ion distribution
becomes strongly dependent on ion concentration for points very close to the
interface, as shown in Fig.\ \ref{Cs1}. This should be contrasted with Fig.\
\ref{pb}, where the ion distribution is independent of bulk concentration.

\section{Experimental Details}\label{Experimntal}

To manipulate ion bulk-concentrations, we used CsI (99.999\%, Sigma Corp.,
Catalog No. 203033) solutions in ultra-pure water, taking advantage of the
$L_3$ resonance of Cs ions at 5.012 keV in anomalous reflectivity measurements.
To control surface charge density, monolayers of dihexadecyl-hydrogen-phosphate
(DHDP, see Fig.\ \ref{Iso1}) (C$_{32}$H$_{67}$O$_{4}$P; MW = 546.86, Sigma
Corp., Catalog No. D2631) were spread from 3:1 chloroform/methanol solutions at
the air-water interface in a thermostated Langmuir trough \cite{Vaknin2003b}.
DHDP was chosen for this study, since it forms a simple in-plane structure at
high enough surface pressures \cite{Gregory1997,Gregory1999} and its
hydrogen-phosphate head-group ({\it R}-PO$_4$H) has a pK$_a$ = 2.1, {\it
presumably} guaranteeing almost complete dissociation [PO$_4^-$]/[{\it
R}-PO$_4$H] $\approx$ 0.99999, with one electron-charge per molecule
($\sigma_0\approx$ 0.4 C/m$^2$).  Monolayers of DHDP were prepared in a
thermostatic, solid Teflon Langmuir trough, and kept under water-saturated
helium environment.  Ultrapure water (NANOpure, Barnstead apparatus;
resistivity, 18.1 M$\Omega$cm) was used for all solution preparations.
Monolayer compression, at a rate of $\sim$ 1 \AA$^2 / $(molecule$\times$min.),
was started 10-15 minutes after spreading to allow solvent evaporation.  The
monolayer was then compressed to the desired surface pressure. During the
compression, the surface pressure was recorded by a microbalance using a
filter-paper Wilhelmy plate. The Langmuir trough is mounted on a motorized
stage that can translate the surface laterally with respect to the incident
beam to allow the examination of different regions of the sample to reproduce
results and monitor radiation damage of the monolayer.

X-ray studies of monolayers at gas/water interfaces were conducted on the Ames
Laboratory Liquid Surface Diffractometer at the Advanced Photon Source (APS),
beam-line 6ID-B (described elsewhere \cite{Vaknin2001}). The highly
monochromatic beam (16.2 and 5.012 keV; $\lambda = 0.765334$ and $\lambda =
2.47374$ {\AA}), selected by a downstream Si double crystal monochromator, is
deflected onto the liquid surface to a desired angle of incidence with respect
to the liquid surface by a second monochromator [Ge(220) and Ge(111) crystals
at 16.2 and 5.012 keV, respectively] located on the diffractometer
\cite{Als-Nielsen1983,Vaknin2001}. Prior to the measurements, the absolute
scale of the x-ray energy was calibrated with six different absorption edges to
better than $\pm$3 eV.

X-ray reflectivity and GIXD are commonly used to determine the monolayer
structure \cite{Als-Nielsen1989,Jacquemain1991,Kjaer1994,Vaknin2001}. Specular
XR experiments yield the electron density profile across the interface, and can
be related to molecular arrangements in the film. Herein, the electron density
profile across the interface is extracted by a two-stage refinement of a
parameterized model that best fits the measured reflectivity by non-linear
least-squares method.  A generalized density profile $\rho(z) =
\rho^{\prime}(z)+i\rho^{\prime\prime}(z)$ of the electron-density (ED) and the
absorption-density (AD) (real and imaginary parts, respectively) is constructed
by a sum of error functions as follows:
\begin{equation}
\rho(z)=\frac{1}{2}\sum_{i=1}^{N+1}\mbox{erfc}\left(\frac{z-z_{i}}{\sqrt{2}\xi_{i}}\right)(\rho_{i}-\rho_{i+1})+\rho_{1}/2,
\label{rho_z}
\end{equation}
where N+1 is the number of interfaces, $\rho_{i} =
\rho_{i}^\prime+i\rho{_i}^{\prime\prime}$; $\rho_{i}^\prime$ and
$\rho{_i}^{\prime\prime}$ are the ED and AD of $i$th slab, $z_{i}$ and
$\xi_{i}$ are the position and roughness of $i$th interface, respectively,
$\rho_{1}$ is the electron density of the solution ($\approx 0.334$ e/\AA$^3$
), and $\rho_{N+2}=0$ is the electron density of the gaseous environment.  The
use of a different roughness $\xi_{i}$ for each interface preserves the
integral of the profile along $Z$ or the electron density per unit area, thus
conserving the chemical content per unit area a. Although small variations are
expected in $\xi_{i}$ for interfaces that separate  rigid portion of a molecule
(hydrocarbon-chains/gas interface and hydrocarbon-chain/headgroup interface,
for instance), somewhat larger variations can occur at different interfaces
(such as, gas/hydrocarbon chains interface versus headgroup/subphase
interface).   The AD profile is particularly important at the Cs resonance
(5.012 keV) as demonstrated below. The reflectivity is calculated by recursive
dynamical methods \cite{Parratt1954,Born1959} of the discretized ED and AD in
Eq.\ (\ref{rho_z}). In the first stage of the refinement, the variable
parameters used to construct the electron density across the interface
$\rho(z)$ are the thickness values of the various slabs $d_{i} =
|z_{i+1}-z_i|$, their corresponding electron densities $\rho_{i}$, and
interfacial roughness $\xi_i$. By non-linear square fit (NLSF) we determine the
minimum number of slabs or `boxes' required for obtaining the best fit to the
measured reflectivity.  The minimum number of slabs is the one for which the
addition of another slab does not improve the quality of the fit, i.e., does
not improve $\chi^2$.  In the second stage, we apply space filling and volume
constraints \cite{Vaknin1991,Vaknin1991b,Gregory1997} to calculate $\rho_i$ by
assigning to each slab a different portion of the molecule, the ions and water
molecules, to a profile that has the same number of slabs as obtained in stage
one of the analysis.  As described in detail in section\ \ref{Results}, the
model eliminates the $\rho_i$'s as parameters, and is self consistent with the
molecular structure, providing detailed information on the constituents of each
slab.

The GIXD measurements are conducted to determine the lateral organization in
the film.  The angle of the incident beam with respect to the surface,
$\alpha$, is fixed below the incident and reflected critical angle
($\alpha_{c}=\lambda(\rho_{s}r_{0}/\pi)^{1/2}; r_{0}=2.82\times10^{-13}$ cm,
where $\rho_{s}$ is the electron density of subphase) for total reflection,
while the diffracted beam is detected at a finite azimuthal in-plane angle,
$2\Theta$, and out-of plane, $\beta$ (the angle of the reflected beam with
respect to the surface). Rod scans along the surface normal at the 2D Bragg
reflections are used to determine the average ordered chain length and tilt
with respect to the surface normal. The intensity along the rod of the 2D Bragg
reflection is analyzed in the framework of the distorted wave Born
approximation (DWBA)
\begin{equation}
I(Q_{xy},Q_{z})\approx|t(k_{z,i})|^{2}|F(Q_{z})|^{2}|t(k_{z,f})|^{2},
\label{NumE}
\end{equation}
where $t(k_{z,i})$ and $t(k_{z,f})$
($k_{z,i}=k_{0}\sin\alpha;k_{z,f}=k_{0}\sin\beta$) are the Fresnel transmission
functions, which give rise to an enhancement at the critical angle. In modeling
the rod scans, the length and tilt of the tails are varied, examining two tilt
directions: one toward nearest neighbors (NN) and the second toward next NN
(NNN)\cite{Kjaer1994,Kaganer1999}. The form factor for the tails is given by
\begin{equation}\label{form}
F(Q_{z}^{'})=\sin(Q_{z}^{'}l/2)/(Q_{z}^{'}l/2)
\end{equation}
where $Q_{z}^{'}$ is defined along the long axis of the tail, and $l$ is the length of the tail.

\section{Experimental Results}\label{Results}

\subsection{Isotherm Comparisons}
\begin{figure}[!]
\includegraphics[width=3.2 in]{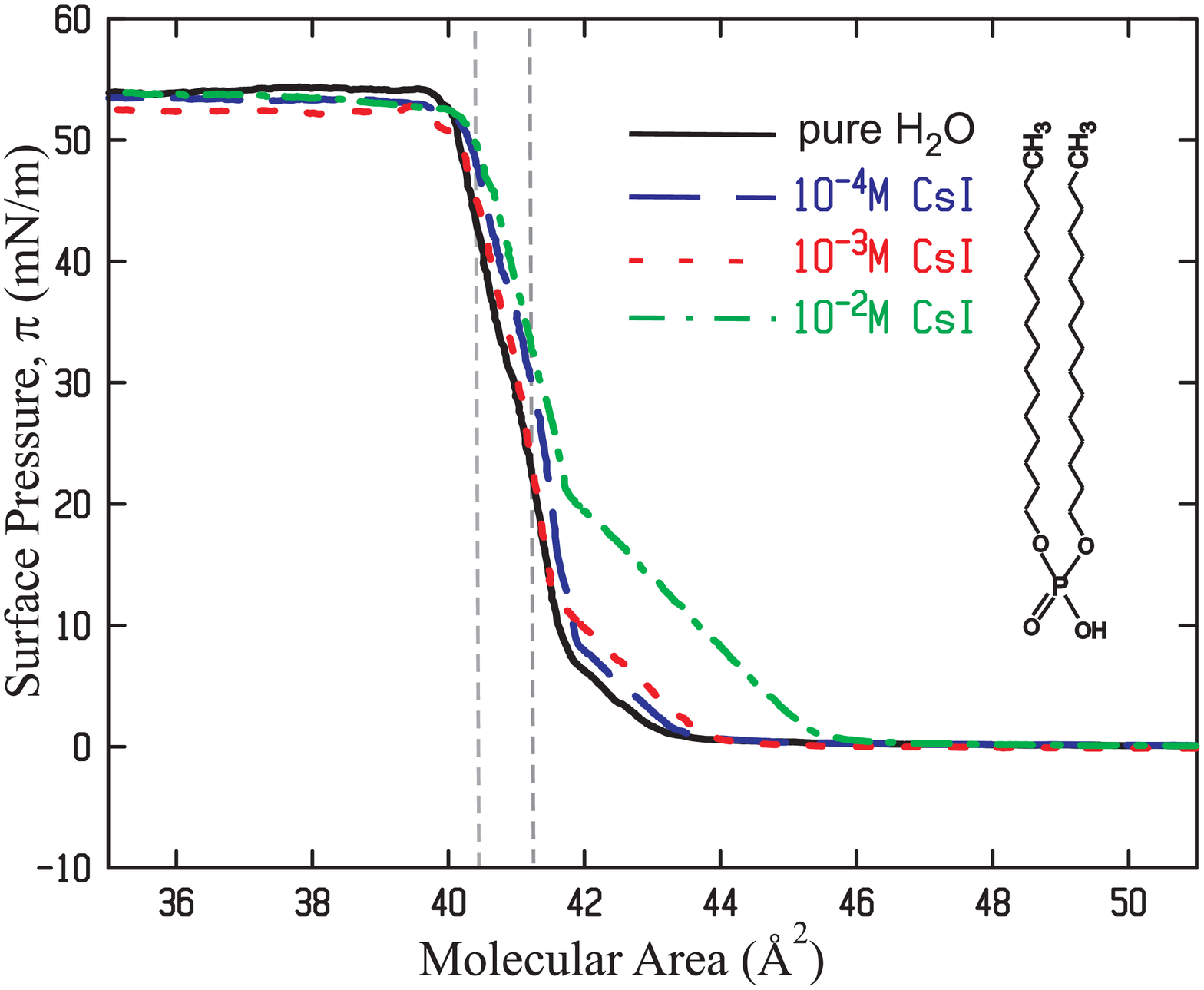}
\caption{\label{Iso1}  Surface pressure versus molecular area for DHDP spread
on CsI solutions at various concentrations as indicated. Reflectivity and GIXD
were performed at constant surface-pressures 30 mN/m and 40 mN/m.}
\end{figure}
Surface pressure versus molecular-area ($\pi-A$) isotherms of DHDP at various
CsI salt concentrations ($n_b$) are shown in Fig.\ \ref{Iso1}. For $\pi > 0$,
the isotherm exhibits two distinct slopes, associated with crystalline tilted
and untilted acyl-chains with respect to the surface normal.  The transition
from tilted to untilted chains at $(A_t,\pi_t)$, occurs at a constant
$A_t\approx 41.5$ {\AA}$^2$, whereas $\pi_t$ increases with salt concentration
$n_b$.  CsI and other electrolytes (NaCl and CsCl) in solution significantly
influence the isotherm causing an increase of the monolayer-coalescence area
$A_C$, (i.e., $\pi > 0$) with the increase in $n_b$.  For $A \lesssim$ 39
{\AA}$^2$, approximately the cross-section of the two acyl-chains of DHDP,
(constant $\pi \approx 55$ mN/m) the monolayer is in the yet poorly
characterized state of {\it collapse}.  In the present study we focus on the
untilted crystalline phase ($30 \lesssim\pi\lesssim 45$ mN/m), where the
molecular area variation at a fixed $\pi$ is less than 1.5\% (in the data
analysis we allow $\pm$ 5\% variation in molecular area).

\subsection{GIXD and Reflectivity off Resonance}
\begin{figure}[!]
\includegraphics[width=3.2 in]{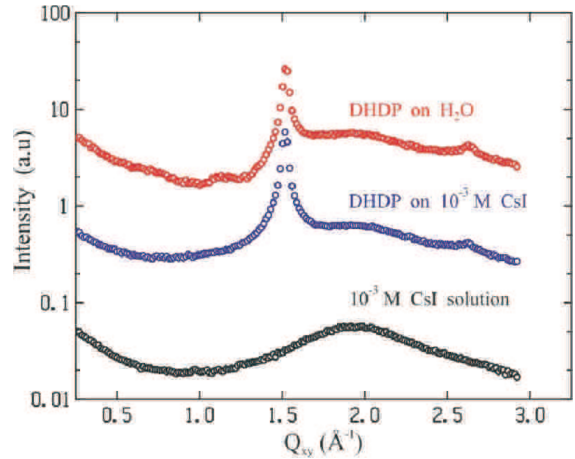}
\caption{\label{gid1} GIXD scans versus the modulus of the inplane momentum
transfer $Q_{XY}$, at surface pressure $\pi=30$ mN/m  (curves are shifted by a
decade for clarity). The Bragg peaks are independent of bulk salt concentration
indicating no significant change in in-plane molecular packing.  GIXD scan for
$10^{-3}$ M CsI (without DHDP) shows a broad peak at $Q_{XY}\approx 2.0$
{\AA}$^{-1}$, due to the surface structure of water.}
\end{figure}

\begin{figure}[!]
\includegraphics[width=3.2 in]{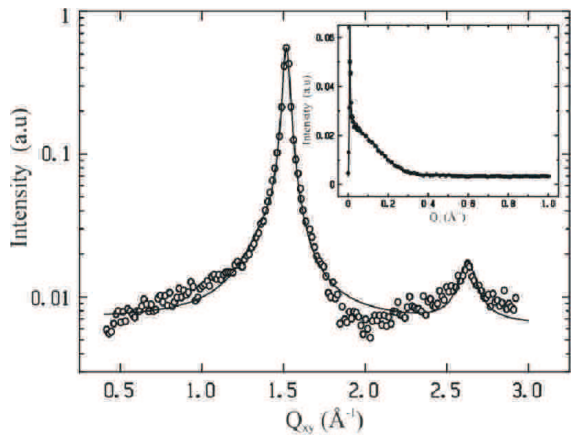}
\caption{\label{gid2} Background subtracted GIXD pattern for a DHDP monolayer
on 10$^{-3}$ M CsI solution ($\pi =30$ mN/m) and the corresponding rod scan
(shown in the inset) at the (1,0) peak (Q$_{XY}=1.516$ {\AA}$^{-1}$). }
\end{figure}

GIXD experiments provided additional insight into the molecular packing of the
acyl chains within the Langmuir monolayers, namely, the average inplane density
of the headgroups and the surface charge-density.  Figure\ \ref{gid1} shows
GIXD scans from DHDP on pure water and on CsI solution (10$^{-3}$ M, at $\pi=
30$ mN/m) and from bare surface of CsI solution (10$^{-3}$ M) before spreading
the monolayer as a function of in-plane momentum transfer $Q_{XY} =
\sqrt{Q_X^2+Q_Y^2}$ \cite{comment1}. The broad peak centered at $Q_{XY} \approx
2$ {\AA}$^{-1}$ is due to the structure-factor of the aqueous-solution
interface.  The spreading and compression of the monolayer slightly modifies
the water structure factor peak at $Q_{XY}\approx 2$ {\AA}$^{-1}$, and brings
about two prominent Bragg reflections due the ordering of acyl-chains
superimposed on a modified surface liquid structure-factor \cite{Vaknin2003b}.
The main features of the diffraction pattern are independent of ionic
concentration, consistent with the isotherms at the 30-40 mN/m region, that
show very small variations in the molecular packing. Figure\ \ref{gid2} shows
the 2D diffraction pattern consists of a strong Bragg reflection at
$Q_{XY}=1.516$ {\AA}$^{-1}$ and a weaker peak at $Q_{XY}=2.627$ {\AA}$^{-1}$,
corresponding to 4.145, and 2.392 {\AA} $d$-spacings, respectively (Table
\ref{gid1_T}).  The shape, spacing, and location of the two intense peaks
correspond closely with literature values for (1,0) and (1,$\bar{1}$) planes in
a hexagonal unit cell of ordered alkyl chains, also confirmed by the ratio
$Q_{XY} (1,\bar{1})/Q_{XY}(1,0) = \sqrt{3}$. This unit cell (molecular area
19.83 {\AA}$^{2}$) agrees with the cross-sectional area of alkyl chain
\cite{Kaganer1999}.  Each headgroup has two alkyl chains, giving a molecular
area of 39.66 {\AA}$^{2}$, in agreement with values obtained from the $\pi - A$
isotherm ($\sim 40.5$ {\AA}$^{2}$, $\pi = 40$ mN/m).  The small variations in
molecular areas are attributed to the existence of domain boundaries, defects
and minute impurities.  The peaks in Fig.\ \ref{gid2} were fitted to
lorentzians the parameters of which are given in Table.\ref{gid1_T}. The peak
linewidth $\Delta$ is significantly larger for the higher order peak. This is
expected in simple 2D crystals \cite{Jancovici1967}, and is even more
pronounced for 2D crystals fluctuating in 3D space (fluctuating tethered
membranes, see Ref. \cite{Bowick2001}).  The inset in Fig. \ref{gid2} shows the
rod-scan at $Q_{XY}$ (1,0) Bragg reflection. The the rod-scan analysis (using
Eqs.\ (\ref{NumE}) and (\ref{form})) yields an average chain-length $\sim 20$
{\AA}, and a tilted angle with respect to the surface normal $ < 5^{\circ}$,
consistent with the reflectivity and previous reports \cite{Gregory1999}.
\begin{table}
\caption{\label{gid1_T}Best-Fit Parameters to high-resolution diffraction scan
of DHDP on $10^{-3}$ M CsI solution ($\pi = 30$ mN/m)}
\begin{ruledtabular}
\begin{tabular}{cccc}
peak & $Q_{XY}$ ({\AA}$^{-1}$) & $\Delta$ ({\AA}$^{-1}$) & Intensity (a.u) \\
(1,0) & 1.516$\pm$0.003&0.026&0.522\\
(1,1) & 2.627$\pm$0.024&0.060&0.009\\
\end{tabular}
\end{ruledtabular}
\end{table}
\begin{figure}[!]
\includegraphics[width=3.2 in]{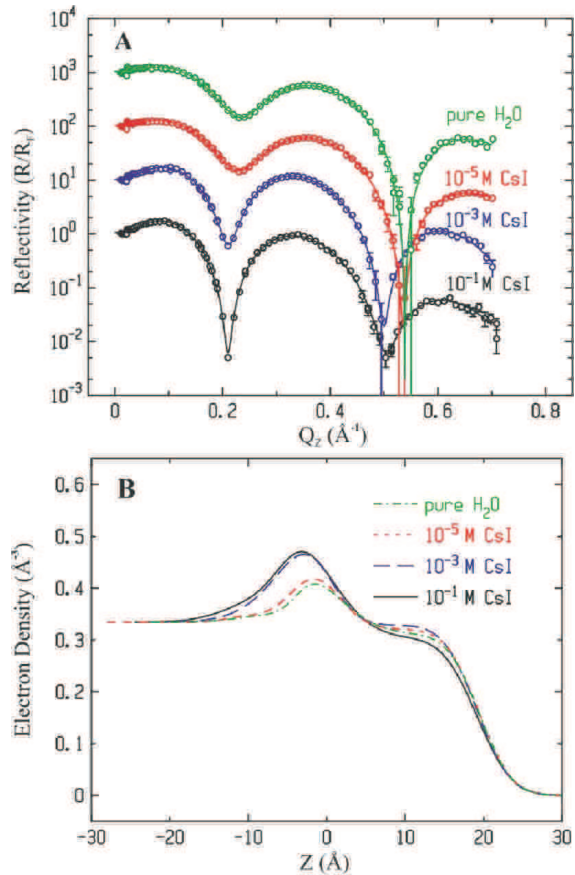}
\caption{\label{ref1} (A) X-ray reflectivity (circles) and corresponding best
fit (solid lines) for the DHDP monolayer at four solutions ($\pi=40$ mN/m)
(curves are shifted by a decade for clarity). (B) ED profiles used to calculate
the fits shown in (A). }
\end{figure}

Figure \ref{ref1}(A) shows the normalized reflectivity curves, $R/R_{F}$ (where
$R_{F}$ is the calculated reflectivity of an ideally flat subphase interface),
for DHDP ($\pi=40$ mN/m) on pure H$_{2}$O and CsI solutions measured at
$E=16.2$ keV. The solid lines are the best-fit calculated reflectivities based
on the ED profiles shown in Fig.\ref{ref1}(B). Similar reflectivity curves were
obtained for $\pi=30$ mN/m.   In Fig.\ \ref{ref1}(A), all x-ray reflectivity
curves differ in the exact position and the sharpness of their minima, and the
intensities of their maxima. Given that, as already shown, the packing of DHDP
is basically independent of salt concentration for $\pi=40$ mN/m, the
reflectivity curves in Fig.\ref{ref1}(A) qualitatively show a strong dependence
of ion distribution close to the interface on bulk ion concentration, in
quantitative agreement with RPB theory as discussed above.
\begin{figure}[!]
\includegraphics[width=3.2 in]{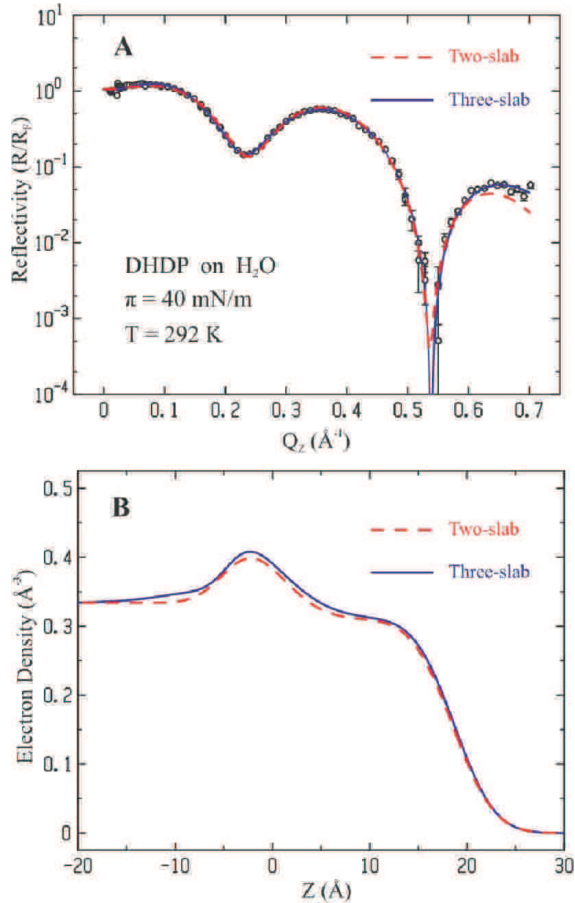}
\caption{\label{ref2} (A) Reflectivity (circles) taken from DHDP on pure H$_2$O
and the best fit by using two-slab model (dashed line) and three-slab model
(solid line). (B) ED profiles extracted from two-slab and three-slab model.}
\end{figure}

 In the first stage of the analysis, we find that the 3-slab ($N = 3$) model provides good
quality fit to all reflectivities and it does not improve with the addition of
more slabs, i.e., more parameters.  Our measured reflectivity for DHDP on pure
H$_{2}$O is consistent with previous measurements \cite{Gregory1997} but
extends to larger momentum transfers ($Q_Z$), allowing for a more refined
structural analysis. Indeed, the two-slab model used in Ref.\
\cite{Gregory1997} was found to be slightly inadequate, particularly at large
$Q_Z$ and a better fit is achieved by adding an extra slab at the
water-headgroup region, as shown in Fig.\ \ref{ref2}. Thus, our detailed
analysis of DHDP on pure water, differs from the one reported in Ref.\
\cite{Gregory1997} in which the headgroup resides on a thin layer (4 - 6 {\AA}
thick) of ED that is just slightly larger than that of bulk water (see Table\
\ref{ref1_T}).  Similar observations  were also reported for other monolayers
at gas/water interface, and were interpreted as interfacial water restructuring
induced by hydrogen bonds.\cite{Lavoie2003} Further evidence of water
restructuring at the interface is also found in the overall GIXD of the
interfacial structure-factor of water, especially a decrease in peak intensity
at $Q_{XY} \approx 2$ {\AA}$^{-1}$ is observed \cite{Vaknin2003b}.
\begin{figure}[!]
\includegraphics[width=3.2 in]{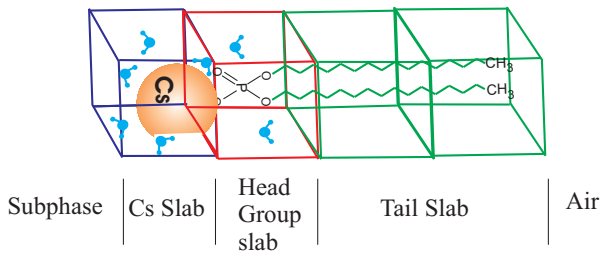}
\caption{\label{stru} Schematic illustration of the three-slab model used to
calculate self consistently the electron-density profile assuming the a DHDP
monolayer of known average molecular area, from GIXD and $\pi - A$ isotherm,
and the associated Cs$^{+}$ and water molecules in the different slabs. Volume
constraints were also applied in the ED calculations of the different potions
of the molecule and the ion-distribution.}
\end{figure}
Modeling DHDP monolayers on the salt solution is slightly more complicated as
Cs$^{+}$ concentration decays slowly as a function of distance from the
interface.  As sketched in Fig.\ \ref{stru}, we assume that Cs ions are present
in both the head-group slab and the slab contiguous to it toward the bulk of
the solution. Table\ \ref{ref1_T} shows the parameters used to produce the ED
profiles in Fig.\ \ref{ref1}(B), and the best-fit shown in Fig.\ \ref{ref1}(A).
The position at $z=0$ is defined by the interface between the phosphate
headgroup and the hydrocarbon chain.  ED profiles show that electron densities
at and below the phosphate headgroup region are higher with the increase of
salt bulk concentration.  The small differences of ED's associated with the
alkyl-chains for the different subphases are due to the minute variations in
molecular areas as shown in the isotherms above.

\begin{table}
\caption{\label{ref1_T} Best-Fit Parameters to the measured reflectivities of
DHDP monolayers at $\pi = 40$ mN/m that generate the ED profiles across the
interface.  In this work, the error estimate (in parentheses) of a parameter is
obtained as described in Refs.\ \cite{Vaknin1991,Vaknin1991b} by fixing a
parameter at different values away from its optimum and readjusting all other
parameters to a new minimum until $\chi^2$ increases by 50\%.  Thicknesses of
head group and Cs box are not well defined due to electron density decay from
$z=0$ to the bulk.}
\begin{ruledtabular}
\begin{tabular}{lllll}
subphase & H$_{2}$O  & $10^{-5}$ M CsI &$10^{-3}$ M CsI & $10^{-1}$ M CsI\\
\hline
$d_{tail}$ ({\AA})&18.7(5)&19.6(10)&19.2(8)&20.2(6)\\
$\rho_{tail}$ ($e$/{\AA}$^{3}$)&0.311(8)&0.320(17)&0.329(14)&0.304(8)\\
$d_{head}$ ({\AA}) &4.4&3.2&3.8&4.3\\
$\rho_{head}$ ($e$/{\AA}$^{3}$)&0.476(15)&0.547(27) &0.590(39)&0.624(38)\\
$d_{Cs}$ ({\AA})&3.6 &9.5 &3.3&4.2\\
$\rho_{Cs}$ ($e$/{\AA}$^{3}$) &0.375(11)&0.347(5)&0.431(14)&0.441(8)\\
\end{tabular}
\end{ruledtabular}
\end{table}
\begin{figure}[!]
\includegraphics[width=3.2 in]{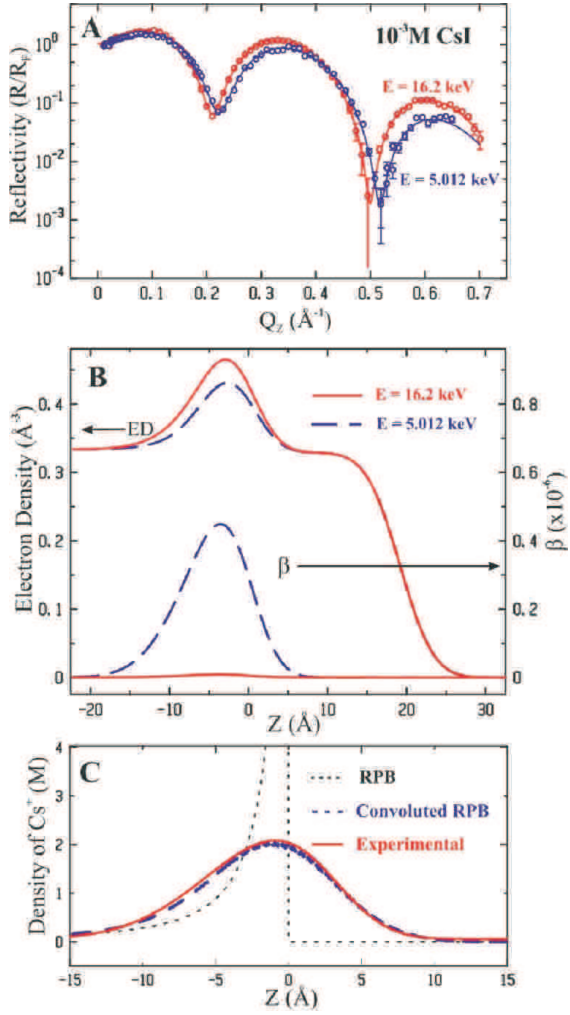}
\caption{\label{ref4} (A) Normalized x-ray reflectivities measured at 16.2 and
5.012 keV for DHDP monolayer spread on 10$^{-3}$ M CsI solution ($\pi=40$
mN/m). The solid lines are calculated reflectivities using the ED profiles
shown in (B). The two data sets were combined and refined to a model with
common structural adjustable parameters. Also shown is the profile of
absorption factor $\beta$, which at 5.012 keV is dominated by the presence of
Cs$^+$ close to the interface. (C) Solid smooth line is the distribution of
Cs$^+$ determined from the reflectivity measurements. The dotted line is the
ion distribution calculated from the RPB equation with the corrected
surface-charge density due to hydronium affinity to PO$_{4}^{-}$. The dashed
line is the RPB result convoluted with a gaussian of width given by the average
surface roughness of the monolayer obtained from XR without any adjustable
parameters.}
\end{figure}
\subsection{Anomalous Reflectivity}
\begin{figure}[!]
\includegraphics[width=3.2 in]{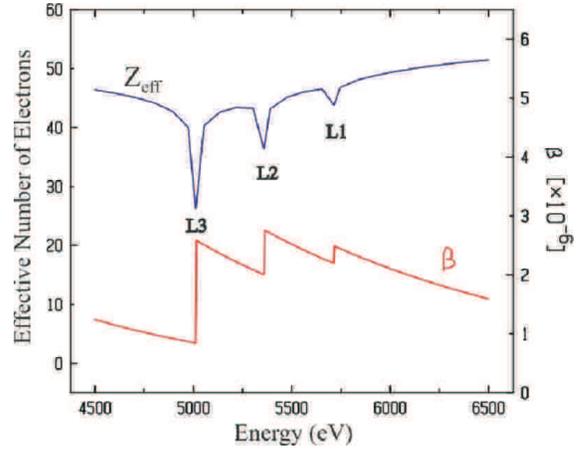}
\caption{\label{zeff} Effective number of electrons and the absorption factor
$\beta$ for Cs$^+$. $Z_{eff}=\rho^\prime/n$, where $n$ is the number density
and $\beta=\lambda^2\rho^{\prime\prime}r_0/2\pi$. }
\end{figure}

Figure\ \ref{ref4}(A) shows reflectivities of DHDP spread on 10$^{-3}$ M CsI
for $\pi$ = 40 mN/m at 16.2 and 5.012 keV.  The reflectivity taken at 16.2 keV
has sharper and deeper minima that are slightly shifted to smaller $Q_Z$,
compared to the reflectivity taken at 5.012 keV.  This is due to the dependence
of $\rho_{i}^{\prime}$ and $\rho_{i}^{\prime\prime}$ on the energy of the x-ray
energy.  At the two energies that the measurements were conducted (16.2 and
5.012 keV), $\rho^\prime$ and $\rho^{\prime\prime}$ dramatically change only
for cesium (as shown in Fig.\ \ref{zeff}) and only slightly for the phosphorous
ion, whereas for the remaining constituents the binding energies are smaller
than 5.012 keV and therefore all electrons behave as free electrons.  Here we
apply stage two of the analysis, each slab is associated with a portion of the
molecule, and the ED's and AD's are calculated self consistently applying
volume constraints.   Thus, the ED of the hydrocarbon slab is given by
\begin{equation}
\rho_{tail}^{\prime} = N_{tail}/d_{tail}A
\end{equation}
and
\begin{equation}
 \rho_{tail}^{\prime\prime} = 0
\end{equation}
where $N_{tail} = 258$ is the total number of electrons in the two acyl chains
\cite{Gregory1997}.   Similarly, we can calculate $\rho_i$ for the headgroup
and for the Cs$^+$ as follows,
\begin{equation}
\rho_{i}^{\prime}=\left(N_{Cs^{+}}Z_{Cs^{+}}+N_{H_{2}O}Z_{H_{2}O}+N_{PO_{4}^{-}}Z_{PO_{4}^{-}}\right)/d_iA,
\label{NumE5}
\end{equation}
\begin{equation}
\rho_{i}^{\prime\prime}=\left(\mu_{Cs}N_{Cs^{+}}/\rho_{0_{Cs}}+\mu_{P}N_{PO_{4}^{-}}/\rho_{0_{P}}\right)/2d_{i}A\lambda
r_{0}, \label{NumE7}
\end{equation}
where $N_j$ is the number of ions or molecules, $Z_j$ is the number of
electrons per ion or molecule, $\mu_j$ is the linear absorption coefficient of
the material when the density of material is $\rho_{0_j}$. $N_{PO_{4}^{-}}$=1
in the headgroup slab and $N_{PO_{4}^{-}}$=0 in the Cs slab (see Fig.\
\ref{stru}).

Using $\rho_{i}^{\prime}$, $\rho_{i}^{\prime\prime}$, and Eq.\ (\ref{rho_z}),
we can create the generalized density $\rho(z) =
\rho^{\prime}(z)+i\rho^{\prime\prime}(z)$, which includes both electron-density
and absorption-density. We then apply the following volume constraints
\begin{equation}
d_{i}A=N_{Cs^{+}}V_{Cs^{+}}+N_{H_{2}O}V_{H_{2}O}+N_{PO_{4}^{-}}V_{PO_{4}^{-}},
\label{NumE6}
\end{equation}
where, $V_{H_2O} = 30$ \AA$^3$, $V_{PO_4^-} = 60$ {\AA}$^3$ (calculated from
the reflectivity of DHDP on water), and $V_{Cs^+} \approx 20$ {\AA}$^3$
(calculated from the ionic radius in standard tables)\cite{comment0}.  The
advantage of this method is that a unique set of parameters is used to fit both
reflectivities at and off resonance simultaneously, thus providing a strong
self-consistency test to the analysis. This is very similar to the approach
developed to determine the structure of a phospholipid monolayer by refining
neutron and x-ray reflectivities simultaneously \cite{Vaknin1991,Vaknin1991b}.
\begin{figure}[htl]
\includegraphics[width=3.2 in]{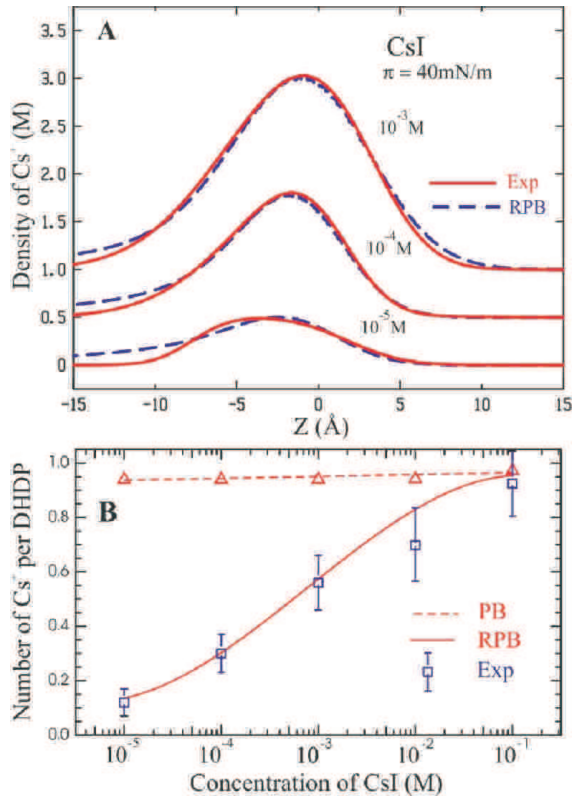}
\caption{\label{Cs1} (A) Interfacial Cs$^+$ distributions (solid lines)
determined from anomalous reflectivities (at 16.2 and 5.012 keV) for various
CsI bulk concentration (shifted by 0.5 M for clarity). Calculated and
convoluted distributions based on RPB with renormalized surface charge density
as described in the text, are shown as dashed lines. (B) Square symbols are
numbers of  Cs$^+$ ions per DHDP ($\approx 41$ {\AA}$^{2}$) by integrating (up
to 15 {\AA}) the experimental distribution obtained from the anomalous
scattering for $10^{-5}$, $10^{-4}$, and $10^{-3}$ M. For $10^{-2 }$ and
$10^{-1}$ M, the integrated number of ions are determined from the
reflectivities off resonance only (the reflectivities at resonance for these
concentrations were not measured).  The dashed line and the solid line are
obtained from PB theory and RPB theory, respectively.}
\end{figure}

The solid lines in Fig.\ \ref{ref4}(A) are calculated from the generalized
density $\rho(z)$, obtained from parameters of a single model-structure for the
combined data sets, as shown in Fig.\ \ref{ref4}(B).  The best fit structural
parameters obtained by this method for various concentrations of CsI in
solution are listed in Table\ \ref{para2}. The absorption factor $\beta$ shown
in Fig. \ref{ref4}(B) can be converted to $\rho^{\prime\prime}$, AD curve, by a
factor ($\beta=\lambda^2\rho^{\prime\prime}r_0/2\pi$).  The AD curve for 5.012
keV up to a normalization factor is practically the profile of the counterion
Cs$^+$ close to the interface (there is a minute contribution to the AD from
phosphorous in the headgroup region, as shown in Eq.\ (\ref{NumE7})). The
difference between the ED's at and off resonance, normalized by $Z_{eff}$(16.2
keV)-$Z_{eff}$(5.012 keV)\cite{Vaknin2003}, gives the desired ionic
distribution at the interface.  Figure\ \ref{ref4}(C) shows (solid line) the
experimental Cs$^+$ distribution close to the interface at 10$^{-3}$ M CsI.
Similar distributions corresponding to other bulk CsI concentrations are shown
in Fig.\ \ref{Cs1}(A) (solid lines).

\begin{table}
\caption{\label{para2} Best-Fit Parameters to the data sets, in which the
reflectivities measured at and off resonance are combined, for various slat
concentrations at $\pi = 40$ mN/m.}
\begin{ruledtabular}
\begin{tabular}{llllll}
subphase(CsI) & $10^{-5}$ M&$10^{-4}$ M&$10^{-3}$ M&$10^{-2}$ M&$10^{-1}$ M\\
\hline

$d_{tail}$ ({\AA})&19.6&19.8&19.2&18.8&20.2\\
$d_{head}$ ({\AA}) &3.2&3.7&3.8&6.7&4.3\\
$N_{Cs^+}$\footnotemark[1] &0.002&0.013&0.270&0.511&0.523\\
$d_{Cs}$ ({\AA})&9.5 &4.7 &3.3&6.9&4.2\\
$N_{Cs^+}$\footnotemark[2]&0.119&0.288&0.289&0.187&0.410\\
\hline
$Area$ ({\AA}$^2$)&41.00&41.04&40.97&41.00&42.08\\
total $N_{Cs^+}$\footnotemark[3]&0.12(5)&0.30(7)&0.56(10)&0.70(12)&0.93(12)\\
\end{tabular}
\end{ruledtabular}
\footnotemark[1] {Number of Cs$^{+}$ in the headgroup slab.} \footnotemark[2]
{Number of Cs$^{+}$ in the Cs slab.} \footnotemark[3] {estimated errors are
given in parentheses.}
\end{table}

\section{Comparison of Experimental Results with Theory}\label{Analysis}

\subsection{Ion Distributions}

We first compute the integrated number of Cs$^+$ per DHDP over the first 15
{\AA} next to the charged interface.  This number can be obtained by
integrating the experimental distribution along the $z$-axis and can be checked
self-consistently from the model used in the analysis of the combined data set
(Table \ref{para2}).  The number of ions per DHDP versus CsI bulk concentration
are plotted in Fig.\ref{Cs1}(B) with square symbols.  In Fig.\ \ref{Cs1}(B) the
integrated values obtained by PB theory with the surface charge corresponding
to the fully deprotonated phosphate groups is also plotted (triangles connected
with a dashed line) for comparison.  As shown, the experimental integrated
number of Cs$^+$ at the interface varies roughly as a power-law of bulk
concentration, which is well described with RPB (solid line) without fitting
parameters, as described in Section~\ref{Theory}.

The ion distribution predicted from RPB theory using Eq.\ (\ref{n_z}) with the
renormalized Gouy-Chapman length (dashed line) is compared with the
experimental distribution in Figs.\ \ref{ref4}(C) and\ \ref{Cs1}(A) (solid
line). As discussed in Section\ \ref{Theory} the theoretical distribution needs
to be convoluted with the effective experimental resolution function.  The
distribution resulting from the convoluted RPB, with no fitting parameters,
with $\Gamma$ ($\Gamma\approx3.8$ {\AA}) obtained from the analysis of the
reflectivity, is shown to reproduce the experimental data remarkably well, as
shown in Fig.\ \ref{ref4}(C).   The value for $pH-pK_a = 4.5$, used in the
calculation, is consistent within the range of our measured values for the $pH
\sim 6.5$.

The distributions corresponding to the three bulk Cs$^+$ concentrations
$10^{-3},10^{-4}$, and $10^{-5}$ M are shown in Fig. \ref{Cs1}(A) with a solid
line. The agreement with the RPB (dashed lines) convoluted as described is
remarkable except for points far from the interface.  We attribute this error
to the difficulty to include slowly decaying tails of the PB theory to the ED
profile modeled by the faster decaying Error functions.

\subsection{Possible deviations from PB theory distributions}

This Section deals with the examination of the sensitivity of the anomalous
reflectivity in determining the ion distribution, in addition to the integrated
number of ions at the interface.  The results presented in Fig.\ \ref{Cs1}(A)
show the unique capability of the anomalous reflectivity technique in providing
ion distributions.  Whereas the integrated number of ions, such as the ones
shown in Fig.\ \ref{Cs1}(B), can be obtained from standard reflectivity
measurements and other experimental techniques, as discussed in Section\
\ref{Introduction}, the determination of ion distributions requires the use of
anomalous reflectivity. In fact, data such as shown in Fig.\ \ref{Cs1}(B) has
been used to assess the validity of PB theory in the past. As discussed in the
Introduction, the excellent agreement maybe somewhat deceptive in that it may
hide significant short-distance deviations from RPB theory because only the
total integrated ion density is involved.
\begin{figure}[!]
\includegraphics[width=3.2 in]{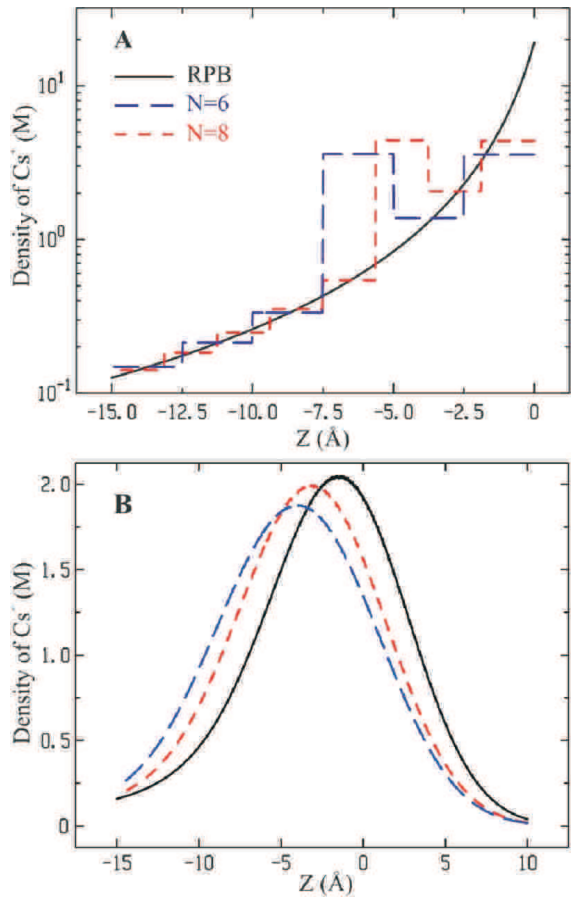}
\caption{\label{Comp5} (A) Solid line corresponds to the RPB profile at bulk
concentration $10^{-3}$ M.  Step-like functions preserve the integral (up the
first 15 {\AA}) of the continuous RPB distribution.  (B) Corresponding
convoluted distributions with a gaussian of width $\Gamma=3.8$ {\AA}.}
\end{figure}
Our experimental resolution constrain possible deviations from RPB
distributions to within 3.8 {\AA}. If such deviations are non-monotonic,
showing bumps or oscillations, then the constraints are even more stringent. As
an example, in Fig.\ \ref{Comp5}(A) we construct two step-wise distributions
($N=6$, and $N=8$ steps of 2.5 and 1.9 {\AA} width, respectively) whose total
integrated area is the same as for RPB (continuous line).  Although the
histograms have the same area as the RPB, they incorporate hypothetical
non-monotonic decay of the distribution in the form of bumps. As it is shown in
Fig.\ \ref{Comp5}(B), despite the fact that the step-size are smaller than the
3.8 {\AA} resolution, such distributions can be ruled out by the experiments.
This demonstrates that if actual ion distributions are non-monotonic, their
maxima or minima must be short-ranged in nature (shorter than $\approx 2$
{\AA}) to be consistent with our data.

\section{Conclusions}\label{Conclusions}
The goal of this study was to explore the accuracy of PB theory for monovalent
ions.  For that, we selected a system with a relatively high surface charge
density (one electron per 40 {\AA}$^2$, lattice constant $\sim 6.8$ {\AA}) and
a low pK$_a = 2.1$ system, which we expected would provide the most favorable
scenario to observe deviations from PB theory.  From the results obtained in
our experiments we conclude that PB with the renormalization of surface charge
density due to proton-transfer and release processes (RPB) is strikingly
accurate.

Certainly, the accuracy of our results is limited by the effective experimental
resolution, which is dominated by the natural surface-roughness of the
gas-amphiphile-water interface ($\approx 3.8$ {\AA}), due mainly to thermally
activated capillary-waves.  We should point out, however, that this resolution
is comparable to the diameter of one Cesium ion ($3.2$ {\AA}) or a single water
molecule ($\sim 3$ {\AA}), and therefore our experimental distributions
constraint deviations from RPB theory to very short-range variations involving
one, at most two, water molecules or cesium ions.  If the actual ionic
distribution is non-monotonic, departures from RPB theory are even more
constrained as it follows from the discussion in Fig.\ \ref{Comp5}.

Our results show that theoretical effects that are usually suggested to modify
PB theory, such as finite ion-size, in-plane modulations of surface charge
density, hydration forces, short-range interactions, the roughness of the
surface or image charges are not necessary to describe the experimental data.
This is not to be understood as implying that such effects are not present, but
rather that their significance is entirely limited to a characteristic spatial
distance of the order of $\approx$ 4{\AA} or less.  As for claims based on the
modification of PB by hydration forces (see for example, Ref.\
\cite{Manciu2004}), our experimental results conclusively rule out the
possibility of modifications of PB within the 10 - 20 {\AA} range from the
interface and confine such corrections, if present, to within the first 3 - 4
{\AA} from the interface as discussed.  Although the experimental counter-ion
distributions are well described within the RPB theory, we point out that the
reflectivity and GIXD hint at water restructuring at the interface.  Future
theoretical or numerical work may clarify this issue.

A recent report analyzing the accuracy of PB near charged liquid-liquid
interfaces by the use of x-ray surface sensitive techniques shows that ion
distributions are well described by PB at concentrations of 10 mM, but marked
deviations are found at higher concentrations (of the order 100 mM), where
ion-ion correlations have to be included in the theoretical
analysis.\cite{Luo2006}  We did not perform anomalous reflectivity for
concentrations above 10mM, so we are not able to provide ionic distributions
for these concentrations. We point out, however, the good agreement found for
the integrated quantities at these concentrations (Fig.\ \ref{Cs1}(B)), which
provides an example of integrated quantities possibly hiding deviations from
actual distributions as pointed out in the introduction.

The results presented in this study enhance our understanding of the
electrostatics in aqueous media, and also show the strength of surface
sensitive x-ray synchrotron techniques in obtaining high resolution data.

{\acknowledgements We thank D. S. Robinson and D. Wermeille for technical
support at the 6-ID beamline. The Midwest Universities Collaborative Access
Team (MUCAT) sector at the APS is supported by the U.S. Department of Energy,
Basic Energy Sciences through the Ames Laboratory under Contract No.
W-7405-Eng-82. Use of the Advanced Photon Source is supported by the U.S.
Department of Energy, Basic Energy Services, Office of Science, under Contract
No. W-31-109-Eng-38.  The work of AT is partially supported by NSF grant
DMR-0426597.}

%GATHER{Soft.bib}
%GATHER{PhysChem.bib}


\begin{references}
\bibitem{Guoy1910} Guoy, A. \emph{J. Phys.} (Paris) {\bf 1910}, {\it 9}, 457.
\bibitem{Chapman1913} Chapman, D. L. \emph{Phil. Mag.} {\bf 1913}, {\it 25}, 475.
\bibitem{Israelachvili2000} Israelachvili, J. \emph{Intermolecular and surface forces},
Academic Press, London, 2000.
\bibitem{Stern1924} Stern, O. \emph{Z.Elektrochem.} {\bf 1924}, {\it 30}, 508.
\bibitem{Debye1923} Debye, P. P.; H{\"u}ckel, F. \emph{Phys. Z.} {\bf 1923}, {\it 24}, 185.
\bibitem{Grahame1947} Grahame, D. C. \emph{Chem. Rev.} {\bf 1947}, {\it 1}, 103.
\bibitem{Outhwaitie1978} Outhwaite, C. \emph{J. C. S. Faraday} {\bf 1978}, {\it 74}, 1214.
\bibitem{Outhwaitie1980} Outhwaite, C.; Bhuiyan, L. B.; Levine, S. \emph{J. C. S. Faraday II} {\bf 1980}, {\it 76}, 1388.
\bibitem{Torrie1982} Torrie, G. M.; Valleau, J. P.; Patey, G. N. \emph{J. Chem. Phys.} {\bf 1982}, {\it 76}, 4615.
\bibitem{Kjellander1984} Kjellander, R.; Marcelja, S. \emph{J. Chem. Phys.} {\bf 1984}, {\it 82}, 2122.
\bibitem{Leikin1993} Leikin, S.; Parsegian, V.; Rau, D. C. \emph{Annu. Rev. Phys. Chem.} {\bf 1993}, {\it 44}, 369.
\bibitem{Manciu2004} Manciu, M.; Ruckenstein E.; \emph{Adv. in Coll. and Int. Sci.} {\bf 2004},
{\it 112}, 109.
\bibitem{Faraudo2004} Faraudo, J.; Bresme F.; \emph{Phys. Rev. Lett.} {\bf 2004}, {\it 92}, 236102.
\bibitem{Proesser2001} Proesser, A. J.; Franses, E. I. \emph{Colloids and Surfaces A} {\bf2001}, {\it 178}, 1.
\bibitem{Hunter1981} Hunter, R. J. \emph{Colloid Science}, Academic Press, London, 1981.
\bibitem{McLaughlin1989} McLaughlin, S. \emph{Annu. Rev. Biophys. Chem.} {\bf 1989}, {\it 18}, {113}.
\bibitem{Tajima1971} Tajima, K. \emph{Bull. Chem. Soc. Jpn.} {\bf 1971}, {\it 44}, 1767.
\bibitem{Tajima1970} Tajima, K.; Muramatsu, K.; Sasaki, M. \emph{Bull. Chem. Soc. Jpn.} {\bf 1970}, {\it 43}, 1991.
\bibitem{Kobayashi1986} Kobayashi, K.; Takaoka, K. \emph{Bull. Chem. Soc. Jpn.} {\bf1986}, {\it 59}, 93.
\bibitem{Kjaer1989} Kjaer, K.; Als-Nielsen, J.; Helm, C.; Tippman-Krayer, P.; M\"{o}hwald, H. {\it
J. Phys. Chem. Phys.} {\bf 1989}, {\it 93}, 3200.
\bibitem{Bloch1990} Bloch, J.M.; Yun W. \emph{Phys. Rev. A} {\bf 1990}, {\it 41} 844.
\bibitem{Bedzyk1990} Bedzyk, M. J.; Bommarito, G. M.;  Caffrey, M.; Penner, T. L.
\emph{Science} {\bf 1990}, {\it 248}, 52.
\bibitem{LeCalvez2001} LeCalvez, E.; Blaudez, D.; Buffeteau, T.; Desbat, B.
\emph{Langmuir} {\bf 2001}, {\it 17}, 670.
\bibitem{Travesset2006} Travesset, A.; Vaknin D. \emph{Europhys. Lett.} {\bf 2006}, {\it 74},  181.
\bibitem{Gur1977} Gur, Y.; Ravina, I.; Babchin, A. J. \emph{J. Colloid Interface Sci.} {\bf 1978}, {\it 64
} 333.
\bibitem{Paunov1996} Paunov, V. N.; Dimova, R. I.; Kralchevsky, P. A.; Broze, G.;
Mehreteab, A. \emph{J. Colloid Interface Sci.} {\bf 1996}, {\it 182 } 239.
\bibitem{Vaknin2003} Vaknin, D.; Kr\"{u}ger, P.; L\"{o}sche, M. \emph{Phys. Rev.
Lett.} {\bf2003}, {\it 90}, 178102.
\bibitem{Bu2005} Bu, W.; Vaknin, D.; Travesset A. \emph{Phys.Rev. E} {\bf 2005}, {\it
72R}, 060501.
\bibitem{Fleck2005} Fleck, C. C.; Netz, R. R.;  \emph{Phys. Rev. Lett.} {\bf
2005} {\it 95} 128101.
\bibitem{Safran1994} Safran, S. \emph{Statistical thermodynamics of surfaces, interfaces, and membranes}, Frontiers in Physics, Perseus Publishing, 1994.
\bibitem{Healy1978} Healy, T. W.; White, L. R. \emph{Adv. In Coll. and Int. Sci.} {\bf1978}, {\it 9}, 303.
\bibitem{Vaknin2003b} Vaknin, D. \emph{J. Am. Chem. Soc.} {\bf 2003}, {\it 125},
1313.
\bibitem{Gregory1997} Gregory, B. W.; Vaknin, D.; Gray, J. D.; Ocko, B. M.; Stroeve,
P.; Cotton, T. M.; Struve, W. S. \emph{J. Phys. Chem. B} {\bf1997}, {\it 101}, 2006.
\bibitem{Gregory1999} Gregory, B. W.; Vaknin, D.; Gray, J. D.; Ocko, B. M.;
Cotton, T. M.; Struve, W. S. \emph{J. Phys. Chem. B} {\bf 1999}, {\it 103},
502.
\bibitem{Vaknin2001} Vaknin, D. in {\it Methods in Materials Research},
edited by Kaufmann, E. N. \emph{et al.}; Wiley, New York, 2001; p 10d.2.1.
\bibitem{Als-Nielsen1983} Als-Nielsen, J.; Pershan, P. S.
\emph{Nucl. Instrum. Methods Phys. Rev.} {\bf1983}, {\it 208}, 545.
\bibitem{Als-Nielsen1989} Als-Nielsen, J.; Kjaer, K. in
\emph{Phase Transitions in Soft Condensed Matter}, edited by Riste, T.;
Sherrington, D.; Plenum, New York, 1989.
\bibitem{Jacquemain1991} Jacquemain, D.; Wolf, S. G.; Leveiller, F.; Deutsch, M.; Kjaer, K.;
Als-Nielsen, J.; Lahav, M.; Leiserowitz, L. \emph{Angew. Chem.} {\bf 1992},
\emph{31}, 130.
\bibitem{Kjaer1994} Kjaer, K. \emph{Physica B} {\bf 1994}, {\it 198}, 100.
\bibitem{Parratt1954} Parratt, L. G. \emph{Phys. Rev.} {\bf1954}, {\it 59}, 359.
\bibitem{Born1959} Born, M.; Wolf, E. \emph{Principles of Optics}, MacMillan, New York, 1959.
\bibitem{Lavoie2003} Lavoie, H.; Blaudez, D.; Vaknin, D.; Desbat, B.; Ocko, B. M.; Salesse, C.
\emph{Biophys. J.} {\bf 2002}, {\it 83}, 3558.
\bibitem{comment0} The 3 density parameters $\rho_i$ are now replaced by 5 parameters
in Eqs. (18) and (20)  ($N_{Cs^+}, N_{H_2O}$,  in the headgroup slab and the
slab below, and the molecular area $A$ which is restricted to $41 \pm 2$
{\AA}$^2$). The two volume constraints in Eq. (22) compensate for the use of
two extra parameters.
\bibitem{Vaknin1991} Vaknin, D.; Kjaer, K.; Als-Nielsen, J.; L\"{o}sche, M. \emph{Biophys. J.}
{\bf 1991}, {\it 59}, 1325.
\bibitem{Vaknin1991b} Vaknin, D.; Kjaer, K.; Als-Nielsen, J.; L\"{o}sche, M. \emph{Makromol.
Chem., Macromol. Symp.} {\bf 1991}, {\it 46}, 383.
\bibitem{Kaganer1999} Kaganer, M. V.; M\"{o}hwald, H.; Dutta, P. \emph{Rev. Mod. Phys.} {\bf1999}, {\it 71}, 779.
\bibitem{comment1} The coordinate system used in this manuscript is such that $Q_Z$ is
normal to the liquid surface, $Q_X$ is parallel to the horizontal (untilted)
incident x-ray beam, and $Q_Y$ is orthogonal to both $Q_X$ and $Q_Z$.  The
hydrocarbon chains form two dimensional poly-crystals giving rise to a
diffraction pattern that depends on the modulus of the in-plane momentum
transfer $Q_{XY} = \sqrt{Q_X^2+Q_Y^2}$ and is practically independent of sample
rotation over the Z-axis.
\bibitem{Jancovici1967} Jancovici, B. \emph{Phys. Rev. Lett.} {\bf 1967}, {\it
19}, 20.
\bibitem{Bowick2001} Bowick, M. J.; Travesset, A. \emph{Phys. Rep.} {\bf 2001},
{\it 344}, 255.
\bibitem{Luo2006} Luo, G. M.; Malkova, S.; Yoon, J.; Schultz, D. G.;  Lin, B. H;
Meron, M.; Benjamin, I.;  Vanysek, P.; Schlossman, M. L.; Science, {\bf 2006},
\emph{311}, 216 (This Report became available to us after the submission of
this manuscript).
\end{references}
\end{document}